\documentclass[preprint,12pt]{elsarticle}

\usepackage{amsmath}
\usepackage{multirow}

\journal{Review of Scientific Instruments}

\biboptions{sort&compress}

\begin{document}

\begin{frontmatter}

\title{Developing the radium measurement system for the water Cherenkov detector of the Jiangmen Underground Neutrino Observatory  }

\renewcommand{\thefootnote}{\fnsymbol{footnote}}
\author{
L.F.~Xie$^{a}$,
J.C.~Liu$^{b}$,
S.K.~Qiu$^{a}$,
C.~Guo$^{b}\footnote{Corresponding author. Tel:~+86-01088236256. E-mail address: guocong@ihep.ac.cn (C.~Guo).}$,
C.G.~Yang$^{b,c}$,
Q.~Tang$^{a}\footnote{Corresponding author. Tel:~+86-13974753537. E-mail address: tangquan528@sina.com (Q.~Tang).}$,
Y.P.~Zhang$^{b}$,
P.~Zhang$^{b}$
}
\address{
${^a}$ School of Nuclear Science and Technology, University of South China, Hengyang, China\\
${^b}$ Key Laboratory of Particle Astrophysics, Institute of High Energy Physics, Chinese Academy of Science, Beijing, China\\
${^c}$ School of Physics, University of Chinese Academy of Science, Beijing, China
}

\begin{abstract}
The Jiangmen Underground Neutrino Observatory is proposed to determine neutrino mass hierarchy using a 20~ktonne liquid scintillator detector. Strict radio-purity requirements have been put forward for all the components of the detector. According to the MC simulation results, the radon dissolved in the water Cherenkov detector should be below 200~mBq/m$^3$. Radium, the progenitor of radon, should also be taken seriously into account. In order to measure the radium concentration in water, a radium measurement system, which consists of a radium extraction system, a radon emanation chamber and a radon concentration measurement system, has been developed. In this paper, the updated radon concentration in gas measurement system with a one-day-measurement sensitivity of $\sim$5~mBq/m$^3$, the detail of the development of the radium concentration in water measurement system with a sensitivity of $\sim$23~mBq/m$^3$ as well as the measurement results of Daya Bay water samples will be presented.
\end{abstract}


\end{frontmatter}


\section{Introduction}
The Jiangmen Underground Neutrino Observatory (JUNO), a multipurpose neutrino experiment, was proposed for neutrino mass hierarchy determination by detecting reactor antineutrinos from nuclear power plants as a primary physics goal~\cite{JUNO}. The excellent energy resolution and large fiducial volume anticipated for the JUNO detector offer exciting opportunities for addressing many important topics in neutrino and astro-physics.

\begin{figure}[htb]
\centering
\includegraphics[height=7.5cm]{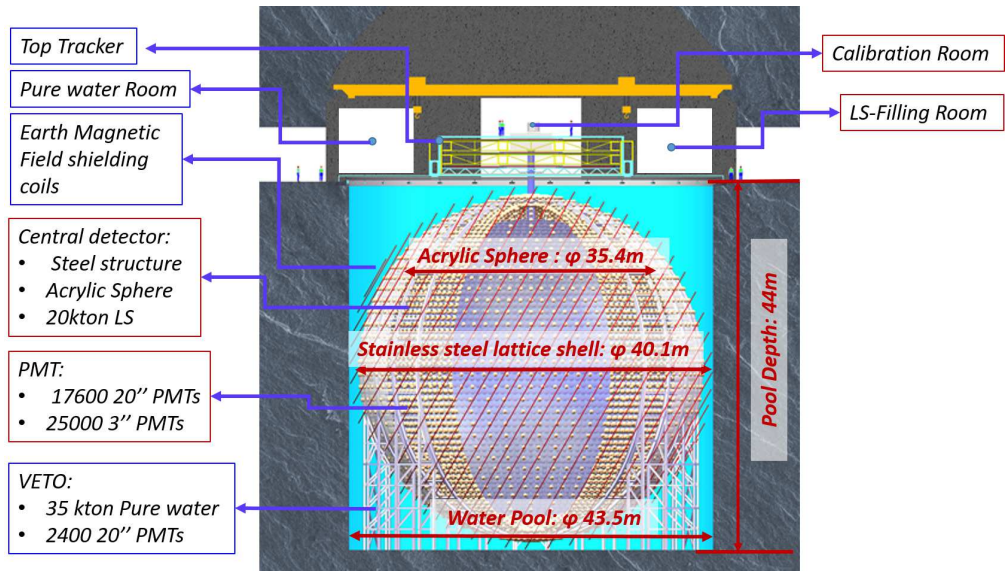}
\caption{Scheme of the JUNO detector. }
\label{fig.JUNO}
\end{figure}

JUNO consists of a central detector (CD) and a veto detector. The central detector, which contains 20 ktonne of liquid scintillator, 17600 of 20 inch photomultiplier tubes (PMT) and 25600 of 3 inch PMTs, is designed to have a very good energy resolution of 3\%@1~MeV and a long operation time of over 20 years. The veto detector, which consists of a water Cherenkov detector and a muon tracker detector, is used for muon detection as well as muon induced background study and reduction. To suppress the radioactive and cosmogenic background, the CD is submerged in the water Cherenkov detector, which is a pool filled with ultra-pure water and instrumented with 2400 of 20 inch microchannel plate photomultiplier tubes (MCP-PMTs). The top tracker (TT) from the OPERA experiment will be placed on the top of the water Cherenkov detector to serve as a cosmic muon tracker.

A reactor electron antineutrino interacts with a proton via the inverse $\beta$-decay (IBD) reaction in the liquid scintillator, resulting in a positron and a neutron. The positron, which carries most of the kinetic energy of the neutrino, deposits its energy quickly, providing a prompt signal. The neutron is captured by a proton after an average time of 200~$\mu$s, then releases a 2.2~MeV $\gamma$, providing a delayed signal. The coincidence of the prompt-delayed signals provides a distinctive antineutrino signature. The estimated IBD reaction rate, after the fiducial cut, energy cut, time cut and vertex cut, in the JUNO detector is $\sim$60/day~\cite{JUNO}. The dominating background is accidentals, coming from two uncorrelated background radiation interactions that randomly satisfy the energy and time correlation for the IBD antineutrino selection. According to the Monte-Carlo (MC) simulation of JUNO, the radon ($^{222}$Rn) concentration in the water Cherenkov detector should be below 200~mBq/m$^3$~\cite{JUNO}. JUNO has very strict radio-purity requirements on all the detector components~\cite{JUNO} and an ultra-pure water system, including the radon removal equipment, will be installed to remove the radon dissolved in water~\cite{RDTM2-48}. $^{226}$Ra, the progenitor of $^{222}$Rn, as one of the radon sources, should also be well controlled.

The method of measuring radium ($^{226}$Ra) content via its gaseous daughter is widely used~\cite{lom2009}. In order to measure the radium concentration precisely, a radium measurement system, which consists of the radium extraction system, the radon emanation chamber and the radon concentration measurement system, has been developed. In this paper, the production of low background manganese fiber (Mn-fiber), the update of the radon concentration measurement system, the calibration of the radium extraction efficiency and the measurement results of radium concentration in water will be presented.

\section{Experimental principle}
\begin{figure}[htb]
\centering
\includegraphics[width=12cm]{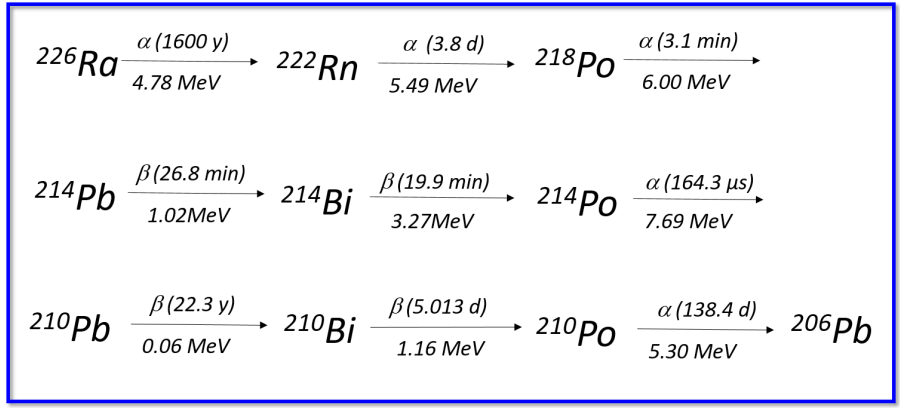}
\caption{The relevant branches of the decay chain of $^{226}$Ra. The time in the bracket is the half-life of the nuclide. The energy for $\alpha$ decay is the energy while for $\beta$ decay is the maximum energy. }
\label{fig.decaychain}
\end{figure}

Figure~\ref{fig.decaychain} shows the relevant branches of the decay chain of $^{226}$Ra~\cite{decay}. The numbers of $^{226}$Ra and $^{222}$Rn atoms at a time $t$ can be expressed according to Eq.~\ref{Eq.Ra} and Eq.~\ref{Eq.Rn}.

\begin{equation}
N_{\rm Ra}(t) = N_{\rm Ra}(0)e^{-\lambda_{\rm Ra} t}
\label{Eq.Ra}
\end{equation}

\begin{equation}
N_{\rm Rn}(t) = \frac{\lambda_{\rm Ra}}{\lambda_{\rm Rn}-\lambda_{\rm Ra}}N_{\rm Ra}(0)(e^{-\lambda_{\rm Ra}t}-e^{-\lambda_{\rm Rn}t})
\label{Eq.Rn}
\end{equation}

Where $N_{\rm Ra}(t)$ is the number of $^{226}$Ra atoms at a time $t$, $N_{\rm Ra}(0)$ is the initial number of $^{226}$Ra atoms, $\lambda_{\rm Ra}$ and $\lambda_{\rm Rn}$ are the decay constants of $^{226}$Ra and $^{222}$Rn, respectively. $N_{\rm Rn}(t)$ is the number of $^{222}$Rn atoms at a time $t$. Considering that $^{226}$Ra has a much longer half life than $^{222}$Rn, namely $\lambda_{\rm Ra} \ll \lambda_{\rm Rn}$, the transfer relation between $^{226}$Ra activity and $^{222}$Rn activity can be simplified to Eq.~\ref{Eq.relation}. Therefore, the activity of $^{226}$Ra can be easily gotten by measuring the emanated $^{222}$Rn activity.

\begin{equation}
A_{\rm Ra}(t) = \frac{A_{\rm Rn}}{1-e^{-\lambda_{\rm Rn}t}}
\label{Eq.relation}
\end{equation}

\section{Experimental setup}
According to Ref.~\cite{1973,1996,SNORa99,1976}, MnO$_{2}$, which has strong adsorbability to radium, can be used to collect the radium ions dissolved in water. Acrylic fiber is widely used as the carrier of MnO$_{2}$~\cite{1973,1996,2010,2018}. The combination of acrylic fiber and MnO$_{2}$ is called Mn-fiber.

During the measurement, water flows through a column that contains Mn-fiber which collects $^{226}$Ra from the flowing water~\cite{1975,1976,1985}. After a certain amount of water passed through the column, the Mn-fiber are drained, removed and then sealed in the Mn-fiber container for radon emanation. The $^{222}$Rn from $^{226}$Ra decays is then swept from the Mn-fiber chamber into an electrostatic chamber where it decays. The charged Po ions from the decays of $^{222}$Rn are carried by an electric field onto a Si-PIN where the $\alpha$s from $^{214}$Po and $^{218}$Po decay can be detected.

In this section, the manufacturing process of the Mn-fiber, the extraction of radium from water as well as the measurement system will be presented.

\subsection{Production of Mn-fibers}
\label{sec.Mnfiber}

\begin{figure}[htb]
\centering
\includegraphics[width=10cm]{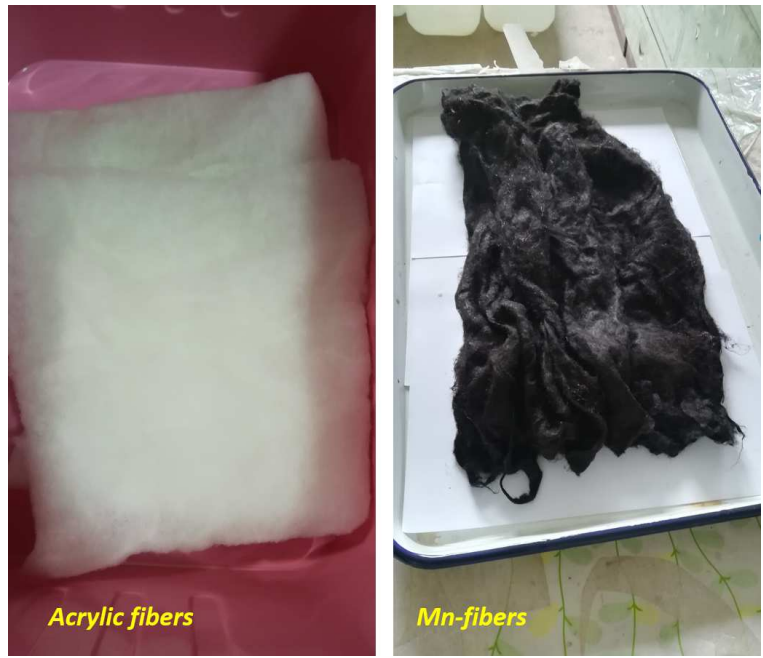}
\caption{Photographs of the acrylic fibers and the Mn-fibers. }
\label{fig.fiber}
\end{figure}

Mn-fibers are made from acrylic fibers, potassium permanganate (KMnO$_4$) solution and concentrated sulfuric acid with mass fraction of 95\%. The production procedure and the ratio of the raw materials are as below:

A. Take some acrylic fibers and wash them repeatedly with ultrasonic and ultra-pure water and dry them completely with a heated vacuum drying oven.

B. Put 50~grams of acrylic fibers into 3~liters of 0.5~mol/L potassium permanganate solution and then pour in 30~ml concentrated sulfuric acid.

C. Heat to boil and keep them boiling for $\sim$2.5~hours, during which the fibers should be properly stirred to make full contact with the solution.

D. Take out the fibers and keep rinsing them with distilled water until the water phase becomes light purple to remove the inefficiently adsorbed MnO$_2$ and the attached KMnO$_4$.

E. Allow the fibers to dry in a clean room. Note that the fibers should not be heated to dry because proper humidity of the fiber is essential to ensure optimum emanation of radon~\cite{sun1998}. In this experiment, the fibers are dried for ~24 hours and the environmental temperature and relative humidity are 21$\pm$1 centigrade and 40\%$\pm$5\% respectively.

F. Keep sealed for use. There is no special rules for sealing and we use aluminium foil bag in this experiment.

The pictures of acrylic fibers and dried Mn-fibers are shown in Fig.~\ref{fig.fiber}.

\subsection{Extraction of radium from water}
\label{sec.extracting}
\begin{figure}[htb]
\centering
\includegraphics[width=10cm]{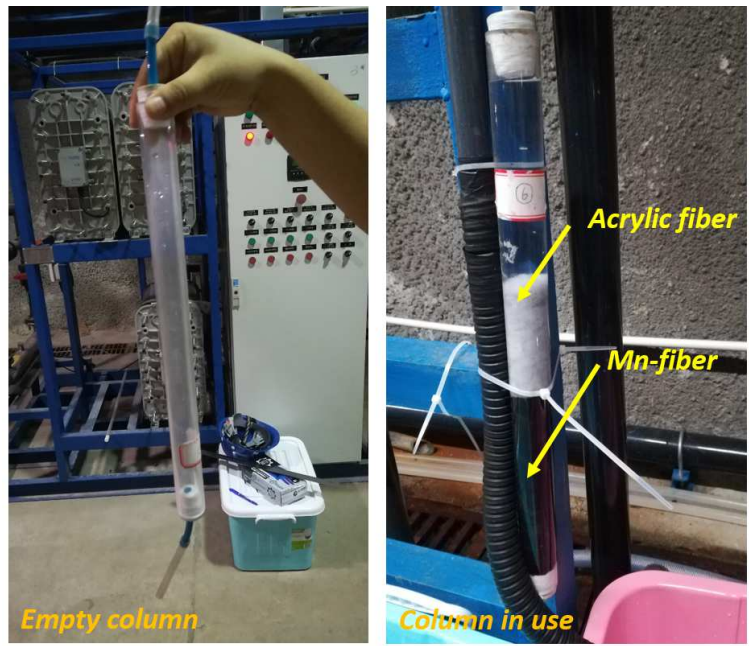}
\caption{Photographs of the empty column and the column loaded with Mn-fibers. }
\label{fig.column}
\end{figure}
The radium dissolved in water will be adsorbed by MnO$_2$  when the water flows over the column loaded with Mn-fibers. Fig.~\ref{fig.column} shows the pictures of the empty column and the column loaded with Mn-fibers. During the radium collection stage, the sealing requirements for the  containers are not strict because the radium content in air can be neglected and the tiny amount of radon adsorbed on the Mn-fibers will be removed at the measurement stage. As the right picture of Fig.~\ref{fig.column} shows, some acrylic fibers are placed at the top of the Mn-fibers to filter out the suspended particulate matters in the water. In order to eliminate the difference in adsorption efficiency caused by different flow velocities, the water flow rate, which is controlled by a float flowmeter (LZB-4WB(F), Senlod Co.Ltd), is set to 250~ml/min during the extraction. Typically, 7.5~g Mn-fiber is used to extract radium from 30~L of water.

\subsection{The radon concentration measurement system}
\label{sec.measurement}

\begin{figure}[htb]
\centering
\includegraphics[height=5cm]{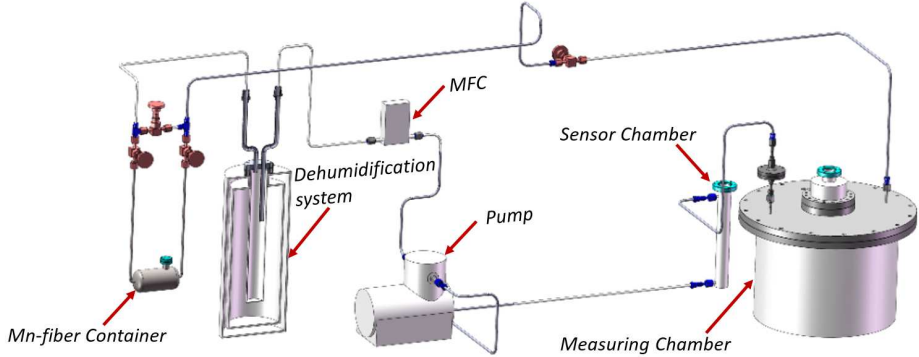}
\caption{Scheme of the system used to measure the radium activity absorbed on the Mn-fibers. Only the dewar and the water container in drawn in this picture, the detail of the dehumidification system is shown in Fig.~\ref{fig.dehumidification}}
\label{fig.mea}
\end{figure}

\begin{figure}[htb]
\centering
\includegraphics[height=5cm]{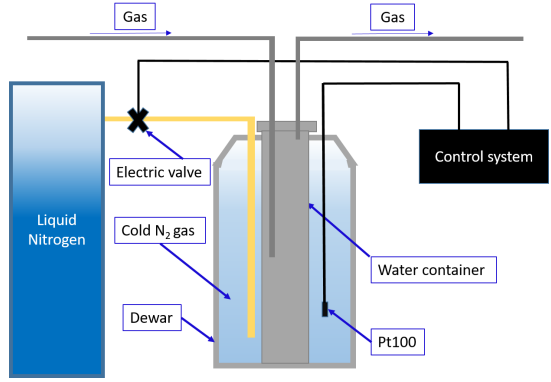}
\caption{Scheme of the dehumidification system. }
\label{fig.dehumidification}
\end{figure}

Fig.~\ref{fig.mea} shows the scheme of the radon measurement system, which is updated based on the radon measurement system developed for the JUNO prototype~\cite{RDTM2-5}. A mass flow controller (MFC, 1\% accuracy, MC-DG901C, Warwick) is used to control the flow rate and a pump (MB-21, Senior) is used for gas circulation. Compared to our previous work, the sensitivity of which was $\sim$9~mBq/m$^3$, the calcium sulfate desiccant has been replaced with the low temperature dehumidification system, which consists of a dewar, a water container and a temperature control system. During the experiment, low temperature nitrogen will intermittently injected into the dewar to keep the inside at $\sim$-50 centigrade. When the gas flows through the water container, heat exchanged between the gas and cryogenic medium, and the water vapor in the gas condenses, freezes and remains in the container. While the radon, which has a much lower boiling point, will directly pass through the system. In our tests, the dehumidification module can keep the relative humidity of the system below 3\%. In this measurement system, knife-edge flanges (CF) with metal gaskets and VCR pipelines with metal gaskets are used and the air leak rate, which is measured by a helium leak detector (ZQJ-3000, KYKY Technology Co. Ltd), is better than 1 $\times$ 10$^{-9}$ Pa*m$^{3}$/s.

The event rate of $^{214}$Po is used to calculate the Rn concentration because there are no other $\alpha$ sources in its signal region as well as $^{214}$Po is known to have a higher collection efficiency than $^{218}$Po~\cite{superk}. In our previous work, the $\sim$9~mBq/m$^3$ sensitivity was derived from the background measurement result, which is (0.389$\pm$0.067) cph (counts per hour). At 90\% confidence level, the sensitivity can be estimated according to Eq.~\ref{Eq.sensitivity}~\cite{superk}.

\begin{equation}
L = \frac{1.64*\sigma_{\rm bg}}{24*C_F}
\label{Eq.sensitivity}
\end{equation}
Where $L$ is the sensitivity in the unit of mBq/m$^3$, $\sigma_{\rm bg}$ is the uncertainty of the background event rate in the unit of cph and $C_F$ is the calibration factor in the unit of cph/(mBq/m$^3$).

\begin{figure}[htb]
\centering
\includegraphics[height=4cm,width=6.5cm]{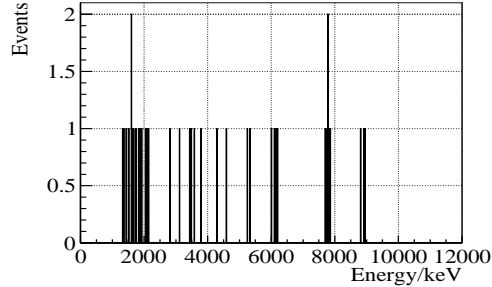}
\caption{The background spectrum for 65 hours measurement. }
\label{fig.bg}
\end{figure}

The background of the former system mainly comes from the desiccant and the residual air inside the detector. While for the updated one, the desiccant has been replaced with the dehumidification system and the system will also be vacuum pumped and flushed with evaporated nitrogen for several times before background measurement, thus the background is lower. The background spectrum, which is taken for 65 hours, is shown in Fig.~\ref{fig.bg}. The counting rate is (0.108$\pm$0.041) cph which results a one-day-measurement sensitivity of $\sim$5~mBq/m$^3$ for radon concentration in gas.

\label{sec.measurement}

In order to ensure the accuracy of radium concentration measurement, strict testing processes have been developed.

A. Take out the Mn-fiber from the column, which is shown in the right part of Fig.~\ref{fig.column}, extract the residual water from the Mn-fibers but keep the Mn-fibers moist because proper humidity is helpful for radon emanation~\cite{sun1998}. Then transfer the Mn-fibers into the Mn-fiber container of the measurement system, which is shown in Fig.~\ref{fig.mea}. The Mn-fiber container is sealed with a CF16 flange.

B. Purge the Mn-fiber container with nitrogen gas for 30 minutes to drive out the residual air and then isolate the Mn-fiber container.

C. Pump the whole system to 100~Pa with a vacuum pump and then fill evaporated nitrogen gas into the system to atmospheric pressure. This process will be repeated three times.

D. Turn on the dehumidification module and circulate the other parts of the system for dehumidification until the relative humidity of the system down to 3\%.

E. Keep the Mn-fiber container sealed for a period of time (at least 24 hours). The $^{226}$Ra, adsorbed by the MnO$_2$, will decay and its daughter, $^{222}$Rn, will be accumulated in the Mn-fiber container during this period.

F. Transfer the gas from the Mn-fiber container to the measurement chamber with a pump and keep circulation for more than 30 minutes so that the gas composition in the system is basically identical.

G. Dehumidify the whole system with the dehumidification module and do not start data taking until the relative humidity decreases to 3\%.

H. Calculate the $^{226}$Ra concentration in water with Eq.~\ref{Eq.CRa}.

\begin{equation}
C_{\rm Ra} = (\frac{(n-n_b)V_s}{C_F(1-e^{-\lambda_{\rm Rn}t})}-mA_{\rm bg})/V_w\varepsilon
\label{Eq.CRa}
\end{equation}

where $C_{\rm Ra}$ is the radium concentration in water in the unit of mBq/m$^3$, $n$ is the event rate of $^{214}$Po for the tested gas in the step G, $n_b$ is the $^{214}$Po event rate of the background, both $n$ and $n_b$ are in the unit of cph, $V_s$ is the volume of the whole system in the unit of m$^3$, $C_F$ is the calibration factor of the radon detector in the unit of cph/(mBq/m$^3$), $\lambda_{\rm Rn}$ is the decay constant of $^{222}$Rn, $t$ is the sealing time of Mn-fiber in the unit of second, $A_{bg}$ is the intrinsic radium background of Mn-fiber in the unit of mBq/g, $m$ is the mass of Mn-fiber in the unit of g, $V_w$ is the volume of water in the unit of m$^3$, $\varepsilon$ is the radium extraction efficiency.

\section{Calibration and measurement results}

In order to get precise results, the intrinsic background and efficiency of the whole system should be well studied. There are two sources of background, namely the background from the radon detector and the intrinsic radium background of the Mn-fiber. The background of Mn-fiber should be measured for each batch because the raw materials and procedures of Mn-fiber production may be slightly different and will result different radium intrinsic background. The detection efficiency of the radon detector will be calibrated with a $^{222}$Rn source and the radium extraction efficiency of Mn-fiber will be calibrated with a known concentration radium solution.

\subsection{Detection efficiency calibration}
\begin{figure}[htb]
\centering
\includegraphics[height=4cm,width=6cm]{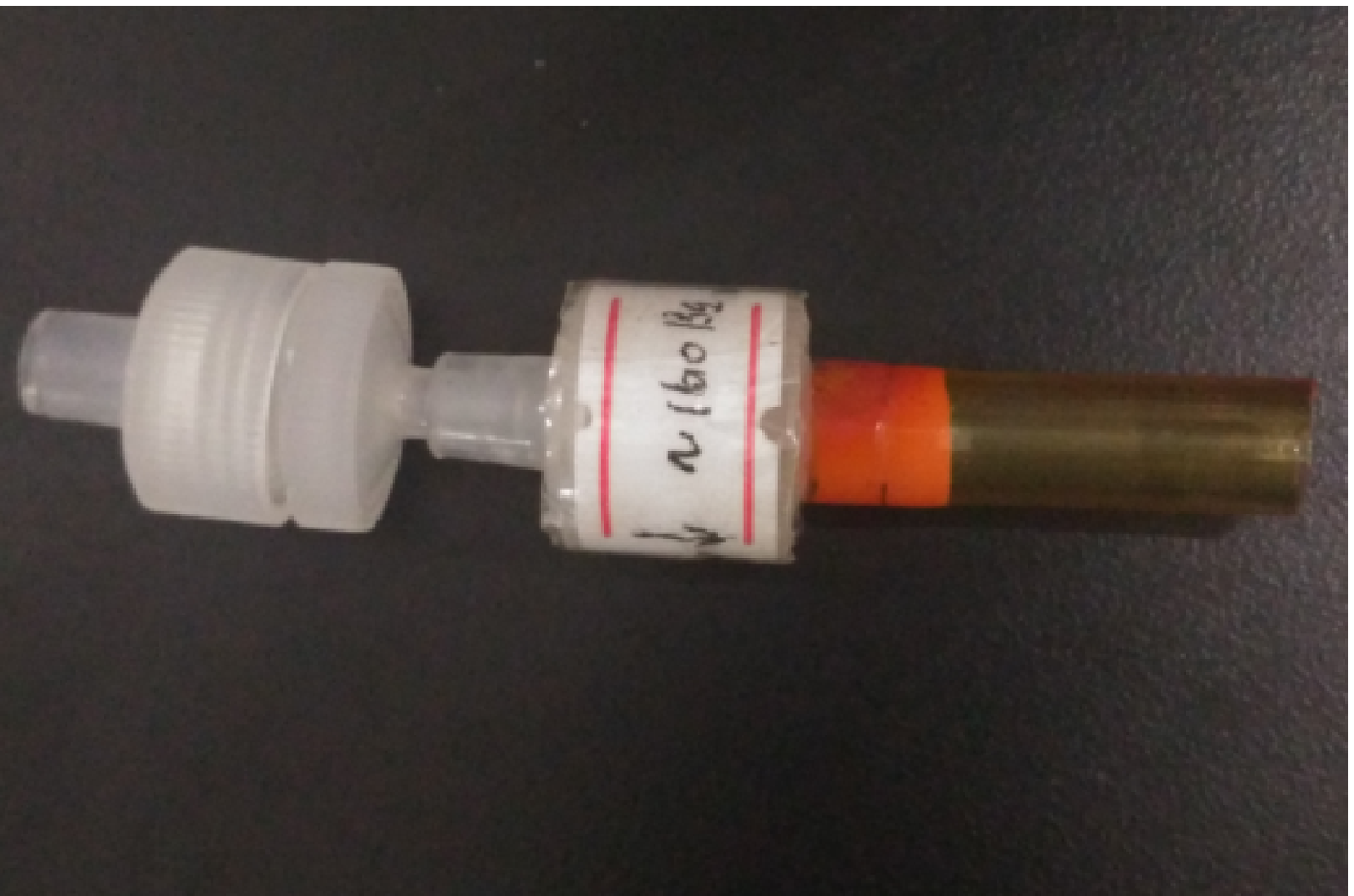}
\includegraphics[height=4cm,width=6cm]{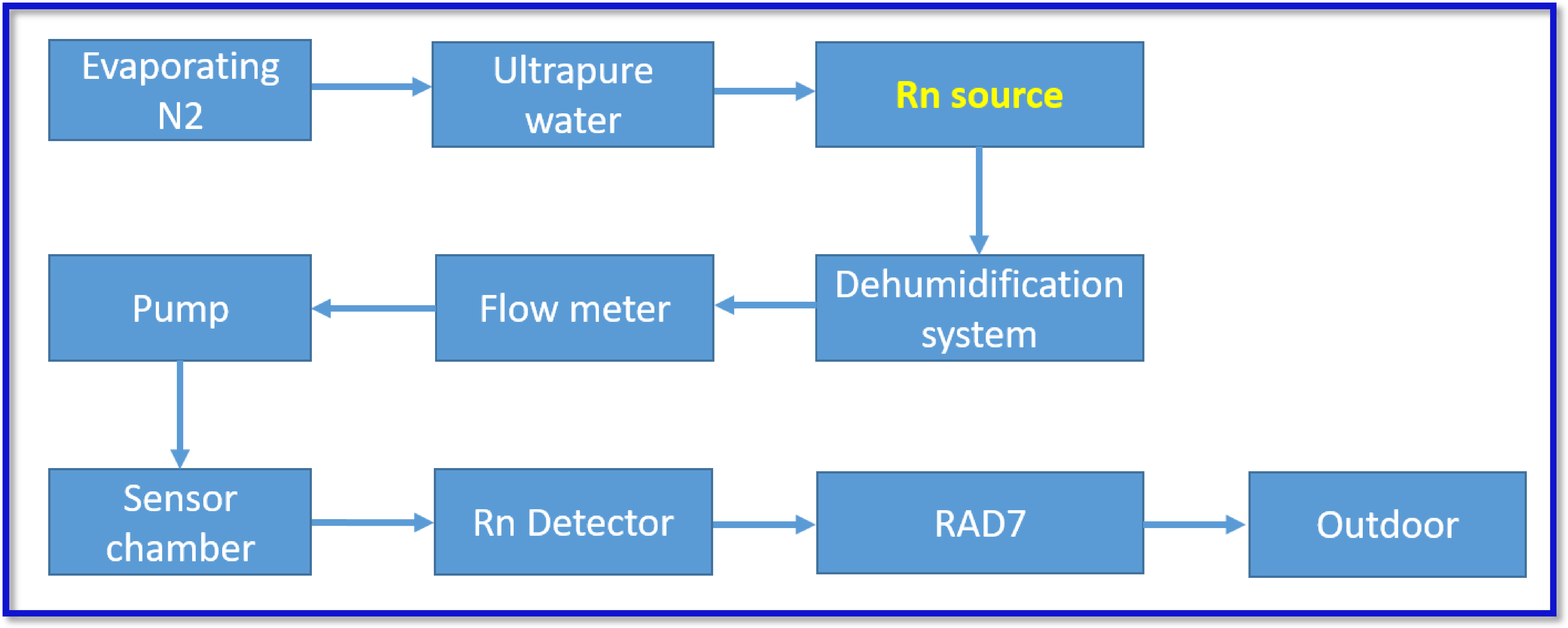}
\caption{Left:The gas flow $^{222}$Rn source. Right: The scheme of Rn detector calibration.}
\label{fig.Rnsource}
\end{figure}

\begin{figure}[htb]
\centering
\includegraphics[width=6.5cm,height=4cm]{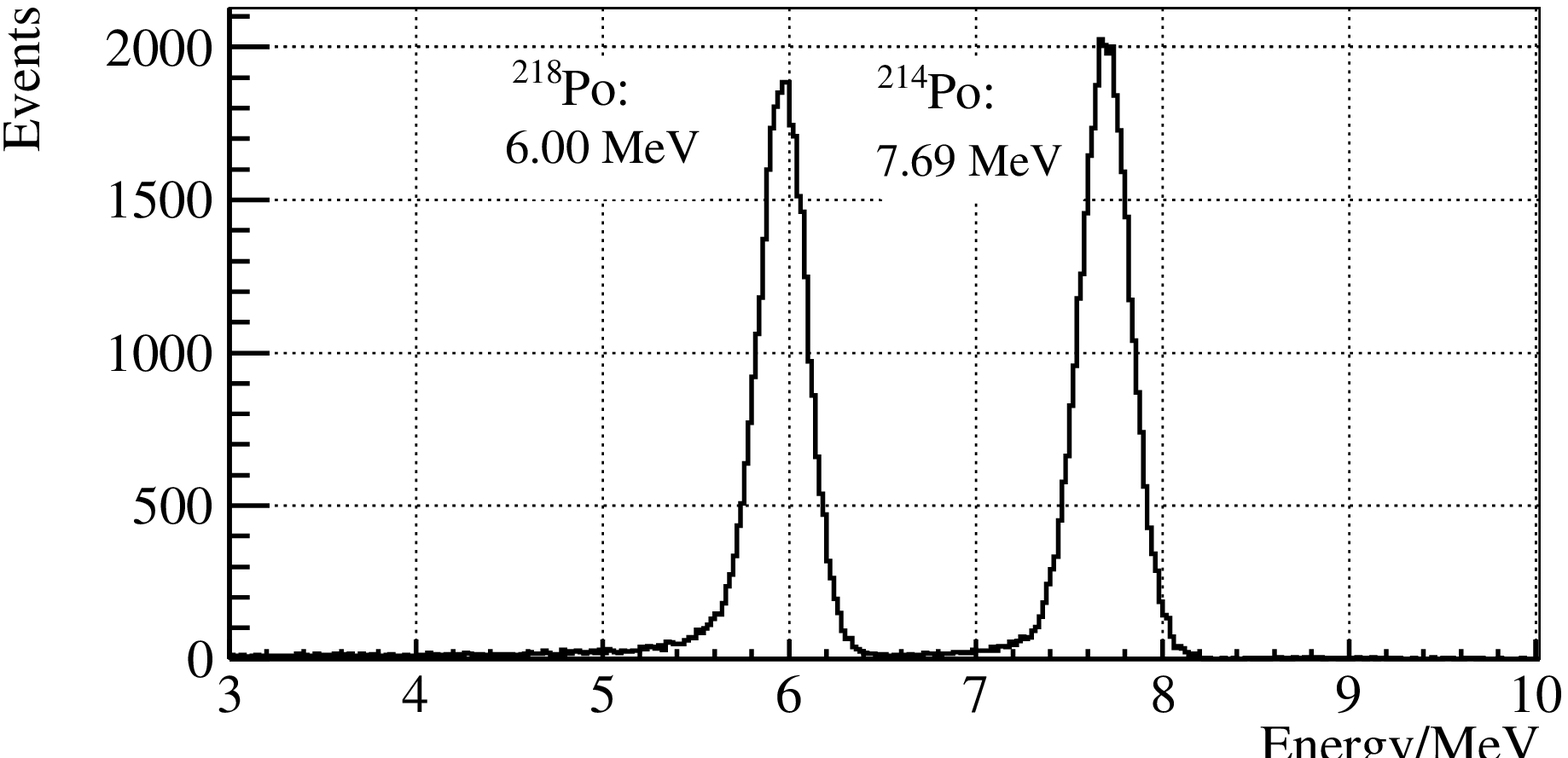}
\includegraphics[width=6.5cm,height=4cm]{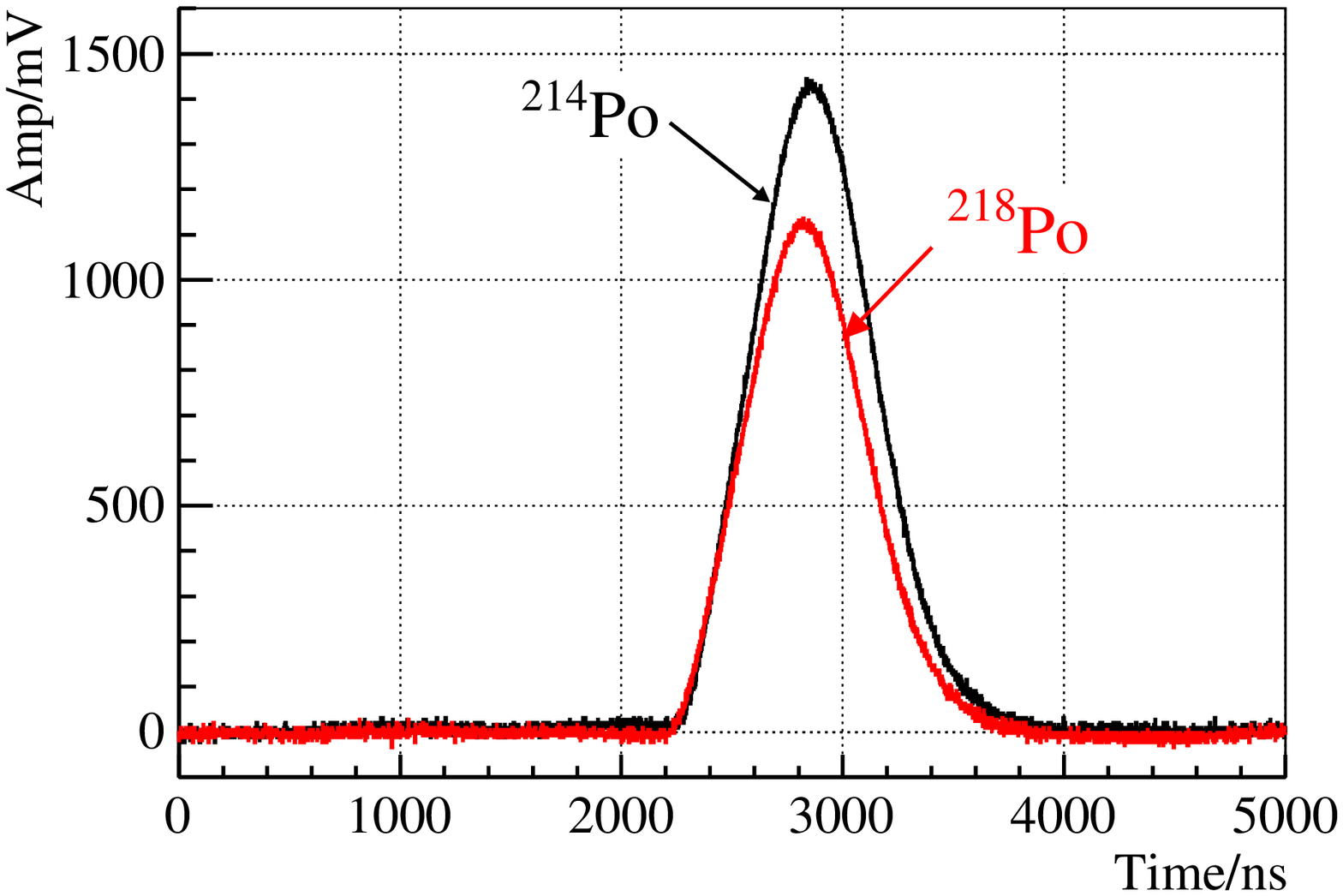}
\caption{Left: Energy spectrum of $^{222}$Rn source. Right: Examples of signal pulses, the black curve is the example pulse of $^{214}$Po and the red curve is the example pulse of $^{218}$Po.}
\label{fig.example}
\end{figure}
The $\alpha$ signals from the $^{222}$Rn decays are recorded with a LeCroy oscilloscope and the detail information of the readout system can be found in Ref.~\cite{RDTM2-5}. The detector is calibrated with a gas flow $^{222}$Rn source, which is made from BaRa(CO$_3$)$_2$ powder by the radon laboratory of South China university. The left part of Fig.~\ref{fig.Rnsource} shows a picture of the radon source and the right part of Fig.~\ref{fig.Rnsource} shows the scheme of using Rn source to calibrate the detector. The gas with a certain humidity is good for radon emanation~\cite{sun1998}, therefore, the gas pass through the ultra pure water before entering the Rn source. Since about 90\% of the radon daughters are positive~\cite{1981} and water vapor is electronegative, the detecting efficiency is relevant to the humidity and it has to be reduced to 3\% before entering into the detector chamber, thus the dehumidification system will be used. During the calibration, the gas flow rate is controlled to be 2~L/min, and the radon concentration in the outgas is 101 $\pm$ 5.9 Bq/m$^3$, which is measured by RAD7 (Durridge company, USA). Fig.~\ref{fig.example} shows example pulses of $^{214}$Po and $^{218}$Po signals as well as the energy spectrum of $^{222}$Rn source. The calibration factor, which characterizes the detection efficiency, can be calculated with the $^{214}$Po event rate and the measured $^{222}$Rn concentration.

\subsection{Radium extraction efficiency calibration}
The radium extraction efficiency should also be calibrated for each batch because the attachments may be oxidized differently, which results two forms of coatings, designated MnO$_2$ and MnO$_x$ (the incomplete oxidation), and the radium adsorption efficiency of them are slightly different~\cite{SNORa03}. Therefore, each batch of the Mn-fibers has to be divided into three parts: for the intrinsic background measurement, for the calibration and for the water sample measurement. The intrinsic background measurement has been carried out in accordance with the steps described in Sec.~\ref{sec.measurement} and the result is (0.158$\pm$0.013) mBq/g for the this batch, the error is statistical only.

\begin{figure}[htb]
\centering
\includegraphics[height=4cm,width=6.5cm]{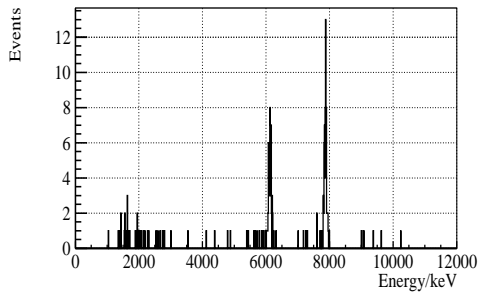}
\caption{Energy spectrum obtained for the $^{226}$Ra solution, the data taking time is 16.22 hours.}
\label{fig.Ra}
\end{figure}

The radium extraction efficiency is calibrated with a known concentration radium solution, which is obtained by diluting 0.5~mL 12.4~Bq/L radium solution (made by Beijing institute of geology of nuclear industry) into 30~L of distilled water. The experiment was carried out in accordance with the steps described in Sec.~\ref{sec.extracting} and Sec.~\ref{sec.measurement}. The energy spectrum of the background and the radium solution obtained by the detector are shown in Fig.~\ref{fig.Ra} and the extraction efficiency, (72.5 $\pm$ 14.6)\%, is derived from the measured $^{214}$Po event rate with Eq.~\ref{Eq.CRa}. The constants (const) in Eq.~\ref{Eq.CRa} are shown in Tab.~\ref{tab:Const} and the values (val) of the variables (var) for calibration in Eq.~\ref{Eq.CRa} are shown in Tab.~\ref{tab:Var_Ra}. The error of $C_{\rm Ra}$ comes from the uncertainty of the radium solution concentration, the error of $C_F$ includes both statistical error and systematic error which is from the uncertainty of radon concentration in the calibration gas, the errors of $n$ and $n_b$ are statistical only and the errors of $t$, $V_s$ and $V_w$ are negligible compared with others.

\begin{table}[htb]
\begin{center}
\caption{Values of the constants used in Eq.~\ref{Eq.CRa}.}
\begin{tabular}[c]{c|ccccccc} \hline
Const & $V_s$~(m$^3$) & $C_F$~(cph/(mBq/m$^3$)) & $\lambda_{\rm Rn}$~(s$^{-1}$) & $m$~(g) & $A_{\rm bg}$~(mBq/g) & $V_w$~(m$^3$) \\\hline
Val & 0.035 & 0.0264$\pm$0.0016 & 2.1$e^{-6}$ & 7.5 & 0.158$\pm$0.013 & 0.03 \\\hline
\end{tabular}
\label{tab:Const}
\end{center}
\end{table}

\begin{table}[htb]
\begin{center}
\caption{Values of the variables used in Eq.~\ref{Eq.CRa} for calibration. }
\begin{tabular}[c]{c|cccc|c} \hline
Var & $C_{\rm Ra}$~(mBq/m$^3$) & $n$~(cph) & $n_b$~(cph) & $t$~(h) & $\varepsilon$ (\%) \ \\\hline
Val & 210.0$\pm$21.0 & 4.32$\pm$0.52 & 0.11$\pm$0.06 & 455.3 & 72.5$\pm$14.6 \\\hline
\end{tabular}
\label{tab:Var_Ra}
\end{center}
\end{table}

\subsection{Sensitivity estimation}
The background of the system, which is from the Mn-fiber and the detector, will be used to estimate the sensitivity of the system. Similar to Eq.~\ref{Eq.sensitivity}, at 90\% confidence level, the sensitivity can be estimated according to Eq.~\ref{Eq.systemsensitivity}:

\begin{equation}
L^{\prime} = \frac{1.64*\sigma^{\prime}_{\rm bg}*V_s}{24*C_F*V_w*\varepsilon}
\label{Eq.systemsensitivity}
\end{equation}

where $V_s$, $C_F$, $V_w$ and $\varepsilon$ have same meanings as Eq.~\ref{Eq.CRa}, $L^{\prime}$ is the sensitivity of the system in the unit of mBq/m$^{3}$ and $\sigma_{\rm bg}^\prime$ is the uncertainty of the total background event rate in the unit of cph. The total background event rate is calculated according to Eq.~\ref{Eq.totalbackground}:

\begin{equation}
n_b^{\prime} = \frac{m*A_{\rm bg}}{V_s}*C_F + n_b
\label{Eq.totalbackground}
\end{equation}

where $n_b\prime$ is the total background event rate in the unit of cph and the other parameters have the same meanings as Eq.~\ref{Eq.CRa}. If 7.5~g Mn-fiber is used to extract the radium from 30~L water, then the sensitivity of the system is $\sim$22.7~mBq/m$^3$ for radium concentration in water measurement. Note that, in this estimation, the sealing time is assumed to be long enough so that $^{222}$Rn has the same activity as $^{226}$Ra and the date taking time is assumed to be 24 hours. The dominating background comes from the Mn-fiber, if its background can be further reduced, the sensitivity will be improved.

\subsection{Measurement results}
The radium concentration measurement system has been applied to the measurement of radium concentration in the water of Daya Bay experiment hall 1 (EH1)~\cite{dybmuon}. The Mn-fibers used in this measurement are in the same batch as those used for background measurement and extraction efficiency calibration, which are described in the previous chapter. The extracting procedures are carried out at the Daya Bay EH1. For the outer pool and inner pool water, the samples are directly pumped out from the outer edges of the inner and outer pools, the sampling point is $\sim$5~m below the water surface. For the water after the water purification system, it is taken from the sampling point of the system. The measurements are carried out at the laboratory of Institute of High Energy Physics (IHEP). The measurement results, shown in Tab.~\ref{tab:dyb}, indicate that the ultra-pure water system at Daya Bay~\cite{dybwater} also has effects on the removal of radium from water. For the Daya Bay experiment, the inner water pool and outer water pool share a set of water circulation and purification system, thus the average radium concentration at the inlet of the water system is (410.2$\pm$105.4)~mBq/m$^3$. Therefore, the radium removal efficiency for the Daya Bay water system is (80.8$\pm$33.6)\%.

\begin{table}[htb]
\begin{center}
\caption{Results of the measurements of the Daya Bay water samples. }
\begin{tabular}[c]{c| c c c |c} \hline
 Location & Outer Pool\\\hline
 Var & $n$~(cph) & $n_b$~(cph) & $t$~(h) & $C_{\rm Ra}$~(mBq/m$^3$) \\\hline
 Val & 1.80$\pm$0.41 & 0.11$\pm$0.04 & 29.5 & 461.2$\pm$99.4 \\\hline
 Location & Inner Pool\\\hline
 Var & $n$~(cph) & $n_b$~(cph) & $t$~(h) & $C_{Ra}$~(mBq/m$^3$) \\\hline
 Val &1.91$\pm$0.24 & 0.11$\pm$0.04 & 40.7 & 359.2$\pm$35.1 \\\hline
 Location&\multicolumn{2}{|c}{After the water purification system} \\\hline
 Var & $n$~(cph) & $n_b$~(cph) & $t$~(h) & $C_{Ra}$~(mBq/m$^3$) \\\hline
 Val & 1.65$\pm$0.44 & 0.11$\pm$0.04 & 161.4 & 78.5$\pm$25.7 \\\hline
\end{tabular}
\label{tab:dyb}
\end{center}
\end{table}

\section{Summary and Prospect}
JUNO has put forward strict requirements on the radioactivity of all the detector components and according to the MC simulation results, the radon concentration in the water Cherenkov detector should be below 200~mBq/m$^3$. $^{226}$Ra, which is the progenitor of $^{222}$Rn and has a half life of 1300 years, should also be taken seriously into account. In order to measure the radium concentration in water, the measurement system, including the radium extracting system, the radon emanation chamber and the radon concentration measurement system, has been developed. Based on the background measurement, the sensitivity of this system is estimated to be $\sim$22.7~mBq/m$^3$ for radium concentration in water measurement. The system has been used for measuring the radium concentration in water at Daya Bay and the radium concentration at different places of EH1 has been presented. The results also show that the ultra-pure water system of Daya Bay could remove $\sim$80\% of the radium from water. For JUNO, the radium removing capability of the ultra-pure water system should also be well studied.

\section{Acknowledgements}
This work is supported by Strategic Priority Research Program of the Chinese Academy of Sciences (Grant No. XDA10010300), National Natural Science Foundation of China (Grant No. 11875280) and the Innovative Project of Institute of High Energy Physics (Grant No. Y9545140U2).

\end{document}